\documentstyle[preprint,aps,prc]{revtex}
\tightenlines
\begin{document}

\title{Center-of-mass effects on the quasi-hole spectroscopic factors in 
the $^{16}$O$(e,e'p)$ reaction}
\author{D. Van Neck and M. Waroquier} 
\address{Department of Subatomic and Radiation Physics, University of Gent, 
Proeftuinstraat 86, B-9000 Gent, Belgium}
\author{A.E.L. Dieperink} 
\address{Kernfysisch Versneller Instituut, 
Zernikelaan 25, 9747 AA Groningen, The Netherlands}
\author{Steven C. Pieper} 
\address{Physics Division, Argonne National Laboratory, Argonne, 
Illinois 60439}
\author{V.R. Pandharipande}
\address{Department of Physics, University of Illinois at Urbana-Champaign, 
1110 West Green Street, Urbana, Illinois 61801}
\maketitle
\vspace*{2cm}
\begin{abstract}
The spectroscopic factors for the low-lying quasi-hole states observed in the
$^{16}$O$(e,e'p)^{15}$N reaction are reinvestigated with a variational 
Monte Carlo calculation for the structure of the initial and final nucleus.
A computational error in a previous report is rectified. It is 
shown that a proper treatment of center-of-mass motion does not lead to a 
reduction of the spectroscopic factor for $p$-shell quasi-hole states, 
but rather to a $7\%$ enhancement. This is in agreement with analytical 
results obtained in the harmonic oscillator model.    
The center-of-mass effect worsens the discrepancy between present theoretical 
models and the experimentally observed single-particle strength. We discuss 
the present status of this problem, including some other mechanisms that may 
be relevant in this respect.
\end{abstract}

\newpage

\section{Introduction}

In present-day $(e,e'p)$ experiments \cite{Kel} one can determine quite 
accurately 
single-particle (s.p.) overlap functions between an $A$-particle nuclear 
target in its ground state and low-lying (quasi-hole) states of the $(A-1)$ 
residual system. The spectroscopic factor, defined as the norm of the overlap
function, is equal to unity only in the fixed-center mean-field 
approximation. Experimentally one finds \cite{Lap} that spectroscopic factors 
of about 
$S\approx 0.6-0.7$ are needed to explain the data.  Such 
deviations of $S$ from unity are normally ascribed to nucleon-nucleon
(NN) correlations. The observed values indicate the sensitivity of the 
quasi-elastic $(e,e'p)$ cross section to correlation effects in the initial 
and final state.  Several nuclear many-body calculations have been made in
attempts to explain the reduced spectroscopic factors of the $p_{1/2}$ 
and $p_{3/2}$ quasi-hole states seen in the $^{16}$O$(e,e'p)$ reaction 
\cite{Rad,Pol,Mut,Geu}.
 
Fixed-center mean-field wave functions are 
not realistic for light systems like $^{16}$O, since they are not 
translationally invariant, i.e.\ they contain spurious center-of-mass 
(c.m.) motion. 
Variational Monte Carlo (VMC) calculations of the $^{16}$O ground state 
\cite{Pie} and the $^{15}$N($p_{3/2}$) quasi-hole state \cite{Rad} , in which 
the nuclear wave functions are explicitly translationally invariant, have 
been made to address this problem.  
It was reported in \cite{Rad} that c.m.\ correlations reduced the $p$-shell
spectroscopic factor by $\sim$ 12\% even in absence of dynamical 
NN correlations.  However, as was noted in \cite{asy}, this 
reduction contradicts the results 
obtained in the harmonic oscillator (h.o.) model, which predicts an 
enhancement by 7\% for the $p$-shell spectroscopic factor in $^{16}$O 
\cite{Die} due to c.m. correlations. 
The discrepancy with the h.o. model values indicated an error in the 
computer program of \cite{Rad}. It has now been identified, and the corrected 
results are presented in the third Section of this paper. 
The correction simply involves a rescaling of the s.p.\ overlap function 
reported in \cite{Rad} by  
the factor $(\frac{16}{15})^{3/2}$. This brings the results without 
dynamical NN correlations into perfect agreement with the h.o.\ results, but 
the full calculation now predicts a $p$-shell spectroscopic factor 
$S_p =0.98$, considerably different from the experimental values.  

The remainder of the paper is organized as follows. In Section II 
we examine the concept of overlap functions with translational invariance of 
the nuclear wave functions taken into account, and point out some 
consequences for the description of knock-out reactions.
We also mention general results of the h.o.\ model 
and apply it to the case of the $^{16}$O nucleus. The theoretical 
 description of the quasi-hole states observed in the $^{16}$O$(e,e'p)$ 
reaction is treated in Section III, where we correct the results of 
\cite{Rad}. It is pointed 
out that a proper treatment of c.m.\ motion in other calculations would also 
worsen the discrepancy with experimental data. In Section IV we 
look at the order of magnitude of two simple corrections to the $(e,e'p)$ 
cross section. The present status of the problem is discussed in Section V. 

\section{Center-of-mass motion and knock-out reactions}
\subsection{Overlap functions in self-bound systems}
For systems which are localized around a fixed force center (e.g.\ the 
electrons of an atom), one defines a s.p.\ overlap function $\phi$ 
between normalized $A$ and $(A-1)$ particle systems as 
\begin{equation}\label{1}
\phi (x_A ) = \sqrt{A}\int dx_1 \dots dx_{A-1} \Phi^{\dagger}_{(A-1)} 
(x_1 ,\dots,x_{A-1})\Phi_{(A)}(x_1 ,\dots, x_A ). 
\end{equation}
(The notation $x_i$ includes the spatial coordinate, ${\bf r}_i$, 
and the appropriate spin and isospin degrees of freedom.)
In the case of Fermi systems, the $\Phi_{(A)}$ and $\Phi_{(A-1)}$ are 
antisymmetric wave functions, and as a consequence, the normalization of 
the overlap function (or spectroscopic factor $S$) has the property 
\begin{equation}
S=\int dx_A |\phi (x_A )|^2 \leq 1.
\end{equation}

In nuclei there is no fixed external force center, but the nucleons are
localized around their c.m.\ due to their mutual interactions.
The eigenstates of such a self-bound system can be factored as
\begin{equation}\label{3}
\Phi_{(A)}(x_1 ,\dots, x_A )=\exp(i{\bf K \cdot R}_A )
\Psi_{(A)}(x_1 ,\dots, x_A ),
\end{equation}
with $\Psi_{(A)}$ the {\em intrinsic}, translationally invariant, wave 
function of the system, and the plane-wave factor describing the c.m.\  
motion; ${\bf K}$ is the total momentum and ${\bf R}_A$ is the position
of the c.m.\ 
The properties of the overlap functions are somewhat different in 
this case.
A detailed analysis of many s.p.\ quantities in self-bound systems 
(the one-body density matrix, natural orbitals, spectral function etc.) 
will be given in 
a future publication \cite{future}.

For the present discussion it will be sufficient to note that, if the intrinsic
wave functions are normalized according to  
\begin{equation}\label{4}
\int dx_1 \dots dx_{A} \delta ({\bf R_A} ) 
|\Psi_{(A)}(x_1 ,\dots, x_A )|^2 = 1 ,
\end{equation} 
then the overlap function $\psi$ must be defined as \cite{del} 
\begin{equation}\label{5}
\psi (x_A )= \sqrt{A}\int dx_1 \dots dx_{A-1} \delta ({\bf R}_{A-1}) 
\Psi^{\dagger}_{(A-1)} 
(x_1 ,\dots,x_{A-1})\Psi_{(A)}(x_1 ,\dots, x_A ).
\end{equation}
As shown in Section II.B, the overlap function defined according
to Eqs.(\ref{4}-\ref{5}) is indeed the counterpart for self-bound systems 
of Eq.(\ref{1}), and it is the natural quantity appearing in the 
description of knock-out reactions when some standard approximations are 
made.  

It should also be kept in mind that the spectroscopic factors in self-bound 
Fermi systems,
\begin{equation}
S=\int dx_A |\psi (x_A )|^2 ,
\end{equation}
can be larger than
one [with deviations of the order ${\cal{O}}(1/A)$]. Although this was already
known before \cite{Die,Cle}, the fact does not seem to be
widely appreciated. The most extreme--though somewhat forced--example is 
that of two identical fermions (e.g.\ two spin-up neutrons). 
The intrinsic wave function can 
be written as $\Psi_{(2)}=f({\bf r}_1 - {\bf r}_2 )$, with
$\int d{\bf r} |f({\bf r})|^2 = 1$. Then the overlap function with the (only) intrinsic
one-particle state, $\Psi_{(1)}({\bf r}_1 )= 1$, is given by 
$\psi ({\bf r}_2 ) = \sqrt{2} f(-{\bf r}_2 )$, and the corresponding spectroscopic 
factor is $S=2$, regardless of the Fermi or Bose nature of the 
particles. A more realistic illustration for three particles is considered 
in the Appendix.

\subsection{Intrinsic transition matrix elements for knock-out reactions}

In this Section we extend the standard derivation of the plane-wave impulse
approximation (PWIA) amplitude in fixed-center systems to the case 
of knock-out reactions on self-bound systems, in order to justify the 
definition Eq.(\ref{5}) for the overlap function.
We consider for simplicity the transition matrix elements of a scalar 
one-particle external probe, e.g.\ the  
operator $\rho ({\bf q})=\sum^A_{j=1} \exp(i{\bf q \cdot r}_j )$, between translationally 
invariant initial and final states of the form (\ref{3}). The matrix 
element is readily separated into a momentum conserving $\delta$-function and 
the intrinsic transition matrix element,
\begin{equation}
\langle \Phi_{(A)f} | \rho({\bf q})| \Phi_{(A)i}\rangle = 
(2\pi)^3 \delta({\bf K}_i +{\bf q} -{\bf K}_f)\langle \Psi_{(A)f} | \rho({\bf q})| \Psi_{(A)i}\rangle ,
\end{equation}
given by 
\begin{equation}
\langle \Psi_{(A)f} | \rho({\bf q})| \Psi_{(A)i}\rangle =
\int dx_1 \dots dx_A \delta ({\bf R}_A) 
\Psi^{\dagger}_{(A)f} (x_1 ,\dots ,x_A ) \sum_{j=1}^A \exp (i{\bf q \cdot r}_j )
\Psi_{(A)i} (x_1 ,\dots ,x_A ).
\end{equation}

In knock-out reactions where the residual $(A-1)$ nucleus is left
in a bound state $\Psi_{(A-1)}$, the intrinsic final state can be specified 
by its asymptotic behavior,
\begin{equation}\label{8}
\lim_{|{\bf r}_A - {\bf R}_{A-1}|\rightarrow \infty}
\Psi_{(A){\bf p}} (x_1 ,\dots ,x_A )=C \exp [i{\bf p} \cdot ({\bf r}_A - {\bf R}_{A-1})]\Psi_{(A-1)}
(x_1 ,\dots ,x_{A-1}),
\end{equation}
with ${\bf p}$ the relative momentum between the knocked-out nucleon 
and the c.m.\ of the remaining ones.

In the plane-wave approximation,  distortion effects on the escaping
particle are neglected, and we simply approximate
\begin{equation}\label{9}
\Psi_{(A){\bf p}} (x_1 ,\dots ,x_A )\approx
\frac{(2\pi)^{-3/2}}{\sqrt{A!(A-1)!}}{\cal{A}}\left\{
\exp [i{\bf p} \cdot ({\bf r}_A - {\bf R}_{A-1})]\Psi_{(A-1)}(x_1 ,\dots ,x_{A-1})\right\} ,
\end{equation}
with $\cal{A}$ the antisymmetrized sum over all coordinate permutations.
The wave functions $\Psi_{(A){\bf p}}$ are not orthogonal to the initial state 
$\Psi_{(A)i}$, and also not mutually orthogonal, since 
\begin{eqnarray}\label{11}
\langle \Psi_{(A){\bf p}}|\Psi_{(A){\bf p}'}\rangle &=&
\int dx_1 \dots dx_A \delta ({\bf R}_A ) \Psi_{(A){\bf p}}^{\dagger}
(x_1 ,\dots ,x_A )
\Psi_{(A){\bf p}'}(x_1 ,\dots ,x_A )\\
&=&\delta ({\bf p}-{\bf p}') - C({\bf p},{\bf p}'),
\end{eqnarray}
with the non-orthogonality correction $C({\bf p},{\bf p}')$ given by
\begin{equation}\label{defc}
C({\bf p},{\bf p}')=(2\pi)^{-3}\int dx dx'\, \rho(x,x')\exp 
[i{\bf r}\cdot({\bf p}'+\frac{1}{A-1}{\bf p})]\exp[-i{\bf r}'\cdot({\bf p}+\frac{1}{A-1}{\bf p}')].
\end{equation}
The quantity $\rho(x,x')$ is the one-body density matrix of the intrinsic
$(A-1)$-particle state $\Psi_{(A-1)}$ \cite{future,del},
\begin{equation}
\rho (x,x') =(A-1) \int dx_1 \dots dx_{A-2} \delta ({\bf R}_{A-2})
\Psi_{(A-1)}^{\dagger} (x_1 ,\dots, x_{A-2}, x)
\Psi_{(A-1)} (x_1 ,\dots, x_{A-2}, x').
\end{equation}
Similarly as for fixed-center systems, the nonorthogonality correction 
$C({\bf p},{\bf p}')$ is suppressed by the momentum dependence of the 
Fourier transform of 
$\rho(x,x')$ for momenta $p$ or $p'$ larger than typical momenta contained 
in the bound-state $\Psi_{(A-1)}$, and can be neglected 
under the usual $(e,e'p)$ kinematical conditions. 

The amplitude itself can now be written as 
\begin{eqnarray}\label{14}
\langle \Psi_{(A){\bf p}} | \rho({\bf q})| \Psi_{(A)i}\rangle &=& \sqrt{A}\sum_{j=1}^A
\int dx_1 \dots dx_A \delta ({\bf R}_A)\ \Psi^{\dagger}_{(A-1)}
(x_1, \dots ,x_{A-1}) \\ \nonumber
&&\exp[-i{\bf p} \cdot ({\bf r}_A -{\bf R}_{A-1})]
\exp[i{\bf q \cdot r}_j]\ \Psi_{(A)} (x_1 ,\dots ,x_A),
\end{eqnarray}
The impulse approximation implies that the nucleon which has absorbed the 
momentum of the probe gets ejected, and corresponds to retaining only the  
term with $j=A$ in Eq.(\ref{14}). The result is 
\begin{equation}\label{14a}
\langle \Psi_{(A){\bf p}} | \rho({\bf q})| \Psi_{(A)i}\rangle\approx 
A_{\mbox{\scriptsize PWIA}}({\bf p},{\bf q}) =
(2\pi)^{-3/2}\int dx_A \exp[-i{\bf r}_A \cdot 
({\bf p}-\frac{A-1}{A}{\bf q})] \psi (x_A) ,
\end{equation}
where $\psi (x_A)$ given by Eq.(\ref{5}), i.e.\ the PWIA amplitude still 
scales with 
the Fourier transform of the overlap function, at missing momentum 
${\bf p} -\frac{A-1}{A}{\bf q}$.

The antisymmetrization correction terms (with $j\neq A$) in Eq.(\ref{14}) 
can be recombined into
\begin{equation}\label{15}
A_{\mbox{\scriptsize A.S.}}({\bf p},{\bf q})=\left( 2\pi \right)^{-3/2} 
\int dx dx' F(x,x') \exp [i{\bf r} \cdot (\frac{A-1}{A}{\bf q}
+\frac{1}{A-1}{\bf p})]\exp [-i{\bf r}' \cdot ({\bf p}
+\frac{{\bf q}}{A})],
\end{equation}
with 
\begin{equation}
F(x,x')=\sqrt{A}(A-1)\int dx_1 \dots dx_{A-2} \delta ({\bf R}_{A-2})
\Psi_{(A-1)}^{\dagger}(x_1, \dots ,x_{A-2},x)
\Psi_{(A)}(x_1, \cdots ,x_{A-2},x,x').
\end{equation}
This correction is again suppressed by the momentum dependence of the Fourier 
transform of $F(x,x')$ when
$p$ and $q$ are large. The magnitude of the nonorthogonality and 
antisymmetrization corrections will be estimated in Section IV.B for the 
kinematical conditions used in the $(e,e'p)$ experiment of \cite{Leu}. 

\subsection{Intrinsic Slater determinants}

Descriptions of all but the lightest nuclei usually involve fixed-center
(shell-model) wave functions of the Slater determinant (SD) type,
\begin{equation} 
\Phi_{(A)}(x_1 , \dots , x_A ) = \frac{1}{\sqrt{A!}}
\mbox{Det}[\phi_{h_j}(x_i)]_{i,j=1,\dots,A}.
\end{equation}
These wave functions contain spurious c.m.\ motion which does not 
correspond to the intrinsic degrees of freedom in self-bound systems. 
The correct intrinsic wave function describing the 
uncorrelated motion of A particles in A orbits around their c.m.\ is given
by
\begin{equation}\label{19}
\Psi_{(A)}(x_1 , \dots , x_A ) = \frac{1}{\sqrt{A!}}
\mbox{Det}[\phi_{h_j}({\bf r}_i -{\bf R}_A, \alpha_i)]_{i,j=1,\dots,A} ,
\end{equation}
where $\alpha_i$ represents spin and isospin degrees of freedom.
These intrinsic Slater determinants (ISD) are obviously translationally
invariant, and should be used in combination with Eqs.(\ref{4}-\ref{5}), 
to calculate s.p.\ overlap functions and spectroscopic factors. 

The effect of the spurious c.m.\ motion on overlap 
functions and spectroscopic factors is non-trivial, even when 
considering transitions between a single ISD $\Psi_{(A)}$ like (\ref{19}) and 
a one-hole state $\Psi_{(A-1)h_n}$ obtained by removing the orbital $h_n$,  
\begin{equation}\label{20}
\Psi_{(A-1)h_n}(x_1 , \dots , x_{A-1} ) = \frac{1}{\sqrt{(A-1)!}}
\mbox{Det}[\phi_{h_j}({\bf r}_i -{\bf R}_{A-1}, \alpha_i)]_{i,j=1,\dots,A; 
i\neq A; j\neq n}.
\end{equation}
In contrast to the situation in fixed-center systems, the presence of the
c.m.\ $\delta$-function in Eqs.(\ref{4}-\ref{5}) makes the calculation of the 
normalization and overlap integrals quite complicated, even for ISD wave 
functions. Only if the orbitals that build up the ISD are chosen to be h.o.\  
wave functions
can the overlap function and spectroscopic factor be evaluated analytically.
The simplest non-trivial illustration is for a h.o.\ well containing three 
neutrons, one spin-down and two spin-up; this is worked out in the 
Appendix. 

The general h.o.\ model has been studied in \cite{Die}, where 
overlaps were considered between an $A$-nucleon ground-state
configuration in the h.o.\  shell model, and the $(A-1)$-nucleon 
one-hole states.
It was found that, in each spin-isospin space $\sigma\tau$, the spectroscopic 
factor for the valence hole state (corresponding to a hole in the 
occupied shell with the largest 
oscillator quantum number $N_v$) is larger than one and given by 
\begin{equation}\label{21}
S^{(\sigma\tau)}_v=\left(\frac{A}{A-1}\right)^{N_v}.
\end{equation}
Moreover, for h.o.\ orbitals the spectroscopic sum rule remains satisfied  
in the
intrinsic frame\footnote{In general, the spectroscopic sum rule is not 
satisfied by considering just one-hole states of ISD's made of non-h.o.\ 
orbitals; this is also illustrated for a three-neutron example in the 
Appendix.}: in each spin-isospin space the sum of the spectroscopic 
factors $S^{(\sigma\tau)}_h$ of all one-hole ISD's yields the number of 
particles $A^{(\sigma\tau)}$
\begin{equation}\label{sumrule}
\sum_h n^{(\sigma\tau)}_h S^{(\sigma\tau)}_h =A^{(\sigma\tau)},
\end{equation}
with $n^{(\sigma\tau)}_h$ the number of particles occupying the main 
h.o.\ shell $h$. Since the sum rule is satisfied and the spectroscopic 
factor for the valence hole state is larger than one, the deeper hole states 
are partly spurious and have $S\leq 1$. 

For the case of $^{16}$O we have $n_s = 1$, $n_p = 3$, $N_v =1$; the 
h.o.\ model predicts for a hole in the $p$-shell 
$S_p =\frac{16}{15}\approx 1.07$, and, through the sum rule (\ref{sumrule}), 
$S_s =4-n_p \times \frac{16}{15} = \frac{4}{5}$ for a hole in the $s$-shell. 
More realistic choices 
of the s.p.\ wave functions, such as Woods-Saxon (W.S.), preclude an 
analytical 
treatment of c.m.\ effects. Nevertheless one can expect results close to 
the h.o.\ values for a light nucleus like $^{16}$O, where h.o.\ and W.S.\  
wave functions are rather similar. We checked this by a direct computation 
of the overlap functions in $^{16}$O with W.S.\ wave functions in the ISD's,  
exploiting the fact that for ISD the many-body integrals in 
Eqs.(\ref{4}-\ref{5}) can 
be reduced to a sequence of one-body integrals. e.g.\ we have (apart from 
normalization factors)
\begin{eqnarray}\label{27}
\psi_{h_n}(x_A)=&\sum&_{m=1}^A \phi_{h_m} (\frac{A-1}{A}{\bf r}_A, \alpha_A )
 \nonumber \\
&\times& \int d{\bf k} \mbox{Det}\left[\int dx \exp[i{\bf k \cdot r}] 
\phi^*_{h_i} (x) \phi_{h_j}
({\bf r}-\frac{1}{A}{\bf r}_A, \alpha )\right]_{i,j=1,\dots ,A; i\neq n; 
j\neq m}.
\end{eqnarray}

As expected the spectroscopic factors obtained with the W.S.\ wave functions
practically coincide with the h.o.\ results.

Note that the correlated many-body wave functions considered in the next 
section are more complicated than single ISD's, and Monte Carlo quadrature 
was used there to calculate the s.p.\ overlap functions and spectroscopic 
factors. We verified that, for the case of single ISD's, the Monte Carlo 
quadrature and a direct evaluation of Eq.(\ref{27}) lead to the same 
result. 

\section{Spectroscopic factors for quasi-hole states in $^{16}$O}

In the variational Monte Carlo framework of \cite{Rad,Pie}, the intrinsic wave 
function of the $^{16}$O ground state has the form 
\begin{equation}
\Psi_{(A)} (x_1 ,\ldots ,x_A) = \hat{F}(x_1,\ldots ,x_A )
\frac{1}{\sqrt{A!}} 
\mbox{Det}\left[ \phi_{h_j} ({\bf r}_i-{\bf R}_A , \alpha_i )
\right]_{i,j=1,\ldots,A}
\end{equation}
The intrinsic 
Slater determinant is built up with the 
s.p.\ wave functions $\phi_h$, and incorporates the mean-field aspects. 
As explained in Section II.C, it is explicitly translationally 
invariant. 
The correlation factor $\hat{F}$ contains two- and three-particle  
correlations of central, spin, isospin, tensor and 
spin-orbit type, and is translationally invariant by itself. 

The correlation factor and the s.p.\ wave functions were determined by 
a minimization of the expectation value of the many-body Hamiltonian
\begin{equation}
H=-\frac{\hbar^2}{2m}\sum_{i=1}^A \nabla^2_i + \sum_{i<j}v_{ij}
+\sum_{i<j<k}V_{ijk},
\end{equation}
where the Argonne $v_{14}$ model of the two-nucleon interaction \cite{Wir} and 
the Urbana model VII of the three-nucleon interaction \cite{Sch} were
used. A detailed description of the variational procedure can be found in 
\cite{Pie}.

The intrinsic wave functions $\Psi_{(A-1)h_n}$ of the quasi-hole states 
in $^{15}$N are approximated as
\begin{equation}\label{27a}
\Psi_{(A-1)h_n} (x_1 ,\ldots ,x_{A-1}) = \hat{F}(x_1,\ldots ,x_{A-1})
\frac{1}{\sqrt{A!}} 
\mbox{Det}\left[ \phi_{h_j} ({\bf r}_i-{\bf R}_{A-1} , \alpha_i )
\right]_{i,j=1,\ldots,A;i\neq A; j\neq n },
\end{equation}
i.e.\ by retaining the correlation factor and s.p.\ orbitals of $^{16}$O and 
omitting in the determinant the column with the corresponding $\phi_{h_j}$. 

The overlap function for the $p_{3/2}$ quasi-hole state 
was then calculated according to Eqs.(\ref{4}-\ref{5}). 
When dynamical NN correlations were neglected (by putting the correlation 
factor $\hat{F} =1$), a spectroscopic factor $S_{p_{3/2}} =0.88$ was found. 
This case
corresponds to the s.p.\ overlap between single intrinsic Slater determinants 
 discussed in Section II.C, for which the harmonic oscillator model predicts
$S_p = 1.07$. The large discrepancy indicates an error in the computer program
used in \cite{Rad}, which has now been identified and corrected. 

To correct the error in \cite{Rad}, 
the s.p.\ overlap functions calculated with translationally 
invariant wave functions simply have to be rescaled by  
$(\frac{16}{15})^{\frac{3}{2}}$. The corrected spectroscopic 
factor with only c.m.\ correlations is thus given by  
$S_{p_{3/2}}=0.88\times (\frac{16}{15})^3 = 1.07$, and is now in perfect 
agreement with the h.o.\ model.  The results including dynamical NN 
correlations in \cite{Rad} must be changed likewise, and we get 
$S_{p_{3/2}}=1.06$ when central NN correlations are added, and 
$S_{p_{3/2}}=0.98$ for the complete calculation. The value 
$S_{p_{3/2}}=0.90$, obtained by including dynamical NN correlations, but no 
c.m.\ correlations, does not change. 

We note that the proper treatment of c.m.\ motion makes the discrepancy 
with the experimentally obtained spectroscopic factors severely worse. 
In \cite{Leu} a spectroscopic factor 
$S_{p_{1/2}}(g.s.)=0.61$ is reported for the $1/2^-$ ground state 
in $^{15}N$, 
whereas the lowest $3/2^-$ state at 6.32 MeV has $S_{p_{3/2}}(6.32)= 0.53$.
The experimental low-lying $p_{3/2}$ strength is fragmented over three 
states at 6.32, 9.93 and 10.70 MeV,  
of which the 6.32 MeV state is the dominant one, with 87\% of the total 
strength. Since we did not include fragmentation due to low-energy 
configuration mixing, our variational result $S_{p_{3/2}}=0.98$ should be 
compared to the total experimental value $S_{p_{3/2}}=0.53/0.87=0.61$.
The c.m.\ correction is not limited to variational 
calculations, but affects other theoretical models as well, such as the 
Green's function calculations in \cite{Pol,Mut,Geu}, for which c.m.\ motion 
was neglected. Thus, the result $S_{p_{3/2}}=0.91$ in \cite{Mut} would 
(in first order) be changed by c.m.\ correlations to 
$S_{p_{3/2}}=0.91\times\frac{16}{15}=0.97$, a value similar to ours,
while that of \cite{Geu} (where low-energy fragmentation is taken into 
account) is increased from $S_{p_{3/2}}(6.32)= 0.76$ to 
$S_{p_{3/2}}(6.32)=0.81$.

\section{Other corrections to the $(e,e'p)$ cross section}

A correct treatment of c.m.\ motion enhances 
the spectroscopic factor of valence hole states at the mean-field level, 
leading e.g.\ to a 7\% enhancement for the $p$-shell spectroscopic factors 
in $^{16}$O. 
This results in present calculations \cite{Rad,Pol,Mut,Geu} of the 
$^{15}$N spectroscopic
factors giving values that are much larger than the experimental values.
In this Section we discuss two simple mechanisms, not taken into account in
\cite{Rad,Pol,Mut,Geu}, that could possibly 
further reduce the theoretical cross section for knock-out of 
a $p$-shell nucleon. Firstly, the spectroscopic factor itself could change 
at the mean-field level if we allow for different mean fields in the 
target and residual nucleus. Secondly, corrections to the PWIA reaction 
amplitude due to nonorthogonality and antisymmetrization may become more
important when c.m.\ motion is taken into account.

\subsection{Different s.p.\ orbitals in target and residual nucleus}

The assumption [implicit in Eq.(\ref{27a})],  of having the same set of 
s.p.\ wave functions describing the target and residual nucleus, is likely 
to become
less adequate as $A$ decreases. In principle one could check this by making 
a separate variational calculation for the $^{15}\mbox{N}$ states, but the 
sensitivity of the results on the shape of the s.p.\ wave functions is not
large enough to do this reliably. In order to have an idea about the 
magnitude  of this effect we calculated $p$-shell spectroscopic factors in a 
fixed-center mean-field model, using slightly different W.S.\ well 
shapes for the protons in $^{15}$N and $^{16}$O, and keeping 
the neutron wells 
identical. In this case it is easy to show that 
$S_p = Q_s^4 Q_p^{10}$, with $Q_s$ ($Q_p$) the overlap between the (slightly) 
different proton $s$-orbitals ($p$-orbitals) in $^{16}$O and $^{15}$N.  
If we fitted the W.S.\ parameters to the experimental charge 
densities available in the literature \cite{deVries}, we find $S_p = 0.946$, 
i.e.\ a substantial $5.4\%$ 
reduction. However, a minimal change in the $^{16}$O geometry (just 
reducing the W.S.\ radius to reproduce the rms radius of $^{15}$N) 
gives only a $1.2\%$ reduction. We tend to regard the latter value as the more 
realistic. The rms radius of $^{15}$N is better established than 
the whole charge density. The $^{15}$N charge density in \cite{deVries} has 
the peculiarity of overshooting the $^{16}$O charge density in the interior, 
and this feature is responsible for the more substantial effect. Also note 
that, lacking information, the neutron wells were taken identical in $^{15}$N 
and $^{16}$O. Relaxing this assumption would also contribute to the reduction 
of the proton $p$-shell  
spectroscopic factor. We estimate therefore the reduction of $S_{p_{3/2}}$ 
due to changes in the s.p.\ wave functions to be at most about 5\% and 
probably less.

\subsection{Nonorthogonality and antisymmetrization corrections}

Here we look at the effects of nonorthogonality 
of the final scattering states $\Psi_{(A){\bf p}}$ in Eq.(\ref{9}) and the 
antisymmetrization 
correction (\ref{15}) of the amplitude. With fixed-center wave functions  
these effects are known to be small under normal $(e,e'p)$ kinematical 
conditions, but it is not inconceivable that c.m.\ effects change this
for intrinsic wave functions. Note e.g.\ that in the fixed-center frame 
the orthogonality corrections vanish if the final state is a SD with one of 
the bound hole states replaced with a continuum state of the same mean field, 
whereas this is no longer the case when ISD are used. 

In order to handle the nonorthogonality corrections on the amplitude, 
we consider the L\"{o}wdin transformation on the set of final scattering 
states $\Psi_{(A){\bf p}}$ \cite{Sch2}. The orthonormal set of L\"{o}wdin 
transformed states can be expressed as 
\begin{equation}
|\tilde{\Psi}_{(A){\bf p}}\rangle = 
\int dp' [N^{-1/2}]_{{\bf pp}'} |\Psi_{(A){\bf p}'}\rangle ,
\end{equation}
in terms of the overlap matrix $[N]$, 
\begin{equation}
[N]_{{\bf pp}'} = \langle \Psi_{(A){\bf p}'} |\Psi_{(A){\bf p}} \rangle = 
\delta ({\bf p}-{\bf p}') -C({\bf p},{\bf p}'),
\end{equation}
given by Eq.(\ref{11}). It can be shown that this transformation does not 
change the correct asymptotic behavior (\ref{8}) of the final scattering 
states. 

The amplitude can now be expanded in powers of the small correction $C$, and 
up to first order we get
\begin{equation}\label{30}
\langle \tilde{\Psi}_{(A){\bf p}} | \rho ({\bf q}) | \Psi_{(A)i} \rangle =
\langle \Psi_{(A){\bf p}} | \rho ({\bf q}) | 
\Psi_{(A)i} \rangle +\frac{1}{2}\int d{\bf p}'
C({\bf p},{\bf p}') \langle \Psi_{(A){\bf p}'} | \rho ({\bf q}) | 
\Psi_{(A)i} \rangle .
\end{equation} 
Using Eqs.(\ref{14a}-\ref{15}) and neglecting antisymmetrization corrections 
in the (small) second term of Eq.(\ref{30}), the amplitude becomes
\begin{equation}\label{31}
\langle \tilde{\Psi}_{(A){\bf p}} | \rho ({\bf q}) | \Psi_{(A)i} \rangle =
A_{\mbox{\scriptsize PWIA}}({\bf p},{\bf q}) + A_{\mbox{\scriptsize A.S.}}
({\bf p},{\bf q}) + A_{\mbox{\scriptsize L\"{o}w}}({\bf p},{\bf q}),
\end{equation}
with the first order L\"{o}wdin correction given by 
\begin{equation}
A_{\mbox{\scriptsize L\"{o}w}}({\bf p},{\bf q}) = -\frac{1}{2}\int d{\bf p}' 
C({\bf p},{\bf p}') A_{\mbox{\scriptsize PWIA}} ({\bf p}',{\bf q}) .
\end{equation}

In the harmonic oscillator model the magnitude of the different terms in 
Eq.(\ref{31}) can be easily estimated by concentrating on the Gaussian part 
of the 
momentum dependence. As an example, the exponent in the Gaussian part of the 
left-hand-side of Eq.(\ref{5}) contains 
\begin{equation}
\sum_{i=1}^A ({\bf r}_i -{\bf R}_A )^2 + \sum_{i=1}^{A-1} 
({\bf r}_i -{\bf R}_{A-1} )^2 = \frac{A-1}{A}({\bf r}_A -{\bf R}_A )^2 
+ 2 \sum_{i=1}^{A-1} ({\bf r}_i -{\bf R}_{A-1} )^2 ,
\end{equation}
and it follows that 
\begin{equation}
\psi (x) = \exp[-\frac{1}{2b^2} \frac{A-1}{A}r^2]\times 
[\mbox{Polynomial in }{\bf r}] .
\end{equation}
As a consequence, the Gaussian part of the momentum dependence of 
$A_{\mbox{\scriptsize PWIA}}$
is  
\begin{equation}\label{35}
A_{\mbox{\scriptsize PWIA}} \sim \exp[-\frac{b^2}{2}\frac{A}{A-1} 
({\bf p}-\frac{A-1}{A}{\bf q})^2].
\end{equation}
Similarly we find 
\begin{eqnarray}
\rho (x,x') &\sim & \exp [-\frac{1}{2b^2}(r^2+{r'}^2)] , \\
C({\bf p},{\bf p}')&\sim & \exp [-\frac{b^2}{2}\frac{A-1}{A-2}
\{({\bf p}'+\frac{1}{A-1}{\bf p})^2+({\bf p}+\frac{1}{A-1}{\bf p}')^2\}] , \\
A_{\mbox{\scriptsize L\"{o}w}}({\bf p},{\bf q})&\sim & \int d{\bf p}' 
C({\bf p},{\bf p}')\exp[-\frac{b^2}{2}\frac{A}{A-1} ({\bf p}'-\frac{A-1}{A}
{\bf q})^2] , \nonumber\\
\label{38}&\sim &\exp[-\frac{b^2}{2}\frac{A}{A-1}
\{({\bf p}+\frac{1}{A}{\bf q})^2+\frac{A-2}{2A}{\bf q}^2\}].
\end{eqnarray}
The last result also holds for the antisymmetrization correction 
$A_{\mbox{\scriptsize A.S.}}({\bf p},{\bf q})$.

The NIKHEF $^{16}$O$(e,e'p)$ experiment \cite{Leu} was performed under 
quasi-elastic 
parallel kinematics, with a roughly constant laboratory proton kinetic 
energy $T_{p_L} \approx 90$ MeV and missing momentum $p_m = p_L - q_L $
scanned in the region (-150 , 250) MeV/c. Under these kinematical 
conditions the magnitude\footnote{Note that in the laboratory frame 
the relative 
momentum $p$ in Section II.B is given by $p=p_L - \frac{1}{A}q_L$} 
of the correction term (\ref{38}) is very small compared 
to the magnitude (\ref{35}) of the leading PWIA term in the amplitude, 
the ratio ranging from $4\times 10^{-6}$ 
at the most negative $p_m$, to $3\times 10^{-5}$ at $p_m$=0, and growing to 
$6\times 10^{-3}$ at the    
most positive $p_m$ (corresponding to the smallest $q_L$). To get a 10\% 
correction at $p_m = 0$, one would need protons ejected with only
$T_{p_L}\approx 20$ MeV. We conclude that, under 
normal $(e,e'p)$ kinematics, 
the nonorthogonality\footnote{There is also the effect of nonorthogonality 
between the scattering states $\Psi_{(A){\bf p}}$ and the initial state 
$\Psi_{(A)i}$, which is straightforward to include in this analysis.
The correction term to the amplitude has a momentum dependence 
$\exp[-\frac{b^2}{2}(\frac{A}{A-1}p^2 +\frac{A-1}{2A}q^2)]$ and is of the  
same order of magnitude as the other correction terms.}
and antisymmetrization corrections are unimportant also 
for the translationally invariant wave functions considered here, 
just as in the fixed-center case.

\section{Discussion}

The observed magnitude of the experimental cross sections of $(e,e'p)$ 
reactions  at small missing energies is not satisfactorily explained 
by present theoretical models. For $^{16}$O we pointed out that c.m.\ 
correlations enhance the cross section leading to $p$-shell quasi-hole
states by about 7\%. 
As a result theoretical predictions \cite{Rad,Pol,Mut} that consider
effects primarily due to short-range and tensor NN correlations
give $S_{p_{3/2}} \approx 0.97$ for $^{15}$N.
Including also low-energy configuration mixing in the target and residual 
nucleus \cite{Geu} provides additional depletion and fragmentation of 
strength and lowers this to $S_{p_{3/2}} \approx 0.81$, 
which is still far above the experimental value 
$S_{p_{3/2}}(6.32) = 0.53\pm 0.05$. At present this
discrepancy does not seem to be understood. 

More generally, the large difference between the experimental value of the 
$p$-shell quasi-hole strength and the present variational result indicates 
that an 
important ingredient is missing in the variational wave function. This is also
signaled by recent calculations for $A\leq 7$ nuclei \cite{Pud}, in which it 
is found that 
the quality of the VMC wave function (used as input in subsequent Green's 
function Monte Carlo calculations) deteriorates with increasing $A$, when 
compared to the final GFMC result. The possibility of $\alpha$-cluster 
components in the surface part of the wave function should be looked at.

\acknowledgements
We thank W.H. Dickhoff for several discussions about c.m.\ effects on
spectroscopic factors. The work of DVN and MW is supported by the Fund for 
Scientific Research-Flanders (FWO), and that of AELD by the foundation for
Fundamental Research of Matter (FOM) of the Netherlands.
The work of VRP is supported by the U. S. National Science Foundation
via grant PHY94-21309, and that
of SCP by the U. S. Department of Energy, Nuclear Physics Division, 
under contract No. W-31-109-ENG-38.
  
\begin{appendix}
\section{Harmonic oscillator model for three particles in a $s^2 p$
configuration}

Consider one spin-down and two spin-up neutrons in a harmonic oscillator well.
This is the simplest non-trivial illustration of the h.o.\ model. 
The ground-state configuration has a spin-up and a spin-down neutron in the 
$s$-shell and one spin-up neutron in the $p$-shell. Its intrinsic wave 
function is
\begin{eqnarray}\label{22}
\Psi_{(3)}({\bf r}_1, {\bf r}_2, {\bf r}_3 ;\mbox{spins})&=& 
\exp[-\frac{b^2}{2}\sum_{j=1}^3 
({\bf r}_j-{\bf R}_3)^2 ]
\left\{ (z_1 - Z_3 )(|uud\rangle-|udu\rangle )\right. \nonumber \\&&\left. 
+(z_2 - Z_3 )(|duu\rangle-|uud\rangle )+
(z_3 - Z_3 )(|udu\rangle-|duu\rangle )\right\},
\end{eqnarray}
where the spin states are differentiated by $u,d$, the occupied $p$-orbital 
is chosen in the $z$-direction, and the h.o.\ length parameter is $b$. 

The three possible one-hole states are
\begin{eqnarray}\label{24}
\Psi_{(2)p}^{(u)} ({\bf r}_1 ,{\bf r}_2, \mbox{spins})&=&
\exp[-\frac{b^2}{2}\sum_{j=1}^2 
({\bf r}_j-{\bf R}_2)^2 ](|ud\rangle -|du\rangle ) , \nonumber\\
\Psi_{(2)s}^{(u)} ({\bf r}_1 ,{\bf r}_2, \mbox{spins})&=&
\exp[-\frac{b^2}{2}\sum_{j=1}^2 
({\bf r}_j-{\bf R}_2)^2 ]((z_1-Z_2)|ud\rangle -(z_2-Z_2)|du\rangle ) , 
\nonumber\\
\Psi_{(2)s}^{(d)} ({\bf r}_1 ,{\bf r}_2, \mbox{spins})&=&
\exp[-\frac{b^2}{2}\sum_{j=1}^2 
({\bf r}_j-{\bf R}_2)^2 ](z_1 - z_2 )|uu\rangle .
\end{eqnarray}

In the notation of Eq.(\ref{21}-\ref{sumrule}) we have 
$n_{s}^{(d)}=n_{s}^{(u)}
=n_{p}^{(u)}=1$, 
$A^{(u)} =2$ and $A^{(d)} = 1$. The spectroscopic factors for 
the valence $p^{(u)}$ and $s^{(d)}$ hole states are, according to 
Eq.(\ref{21}), $S_{p}^{(u)}=\frac{3}{2}$ and 
$S_{s}^{(d)}=(\frac{3}{2})^0 = 1$. The spectroscopic factor for the $s^{(u)}$ 
hole state can be found from the sum rule (\ref{sumrule}),
$S_{s}^{(u)} = 2-\frac{3}{2}=\frac{1}{2}$.

The same results are easily found by applying Eqs.(\ref{4}-\ref{5}) with the 
intrinsic wave functions (\ref{22}-\ref{24}). The normalizations of the 
wave functions are 
\begin{eqnarray}
\langle \Psi_{(3)}|\Psi_{(3)}\rangle &=& \frac{\pi^3 3^{5/2}}{b^8},\\
\langle \Psi_{(2)p}^{(u)}|\Psi_{(2)p}^{(u)}\rangle &=& \frac{\pi^{3/2} 2^{5/2}}
{b^3},\\
\langle \Psi_{(2)s}^{(u)}|\Psi_{(2)s}^{(u)}\rangle &=& \frac{\pi^{3/2} 2^{1/2}}
{b^5},\\
\langle \Psi_{(2)s}^{(d)}|\Psi_{(2)s}^{(d)}\rangle &=& \frac{\pi^{3/2} 2^{3/2}}
{b^5}.
\end{eqnarray}
The overlap functions are then given by 
\begin{eqnarray}
\psi_{p}^{(u)}({\bf r_3})&=& \frac{b^{5/2}2^{5/4}}{\pi^{3/4}3^{3/4}}
z_3 exp[-\frac{b^2}{3}r^2_3],\\
\psi_{s}^{(u)}({\bf r_3})&=& \frac{b^{3/2}2^{1/4}}{\pi^{3/4}3^{3/4}}
exp[-\frac{b^2}{3}r^2_3],\\
\psi_{s}^{(d)}({\bf r_3})&=& \frac{b^{3/2}2^{3/4}}{\pi^{3/4}3^{3/4}}
exp[-\frac{b^2}{3}r^2_3],
\end{eqnarray}
and their normalization agrees with the values for the spectroscopic factors 
mentioned above. 

The spectroscopic sum rule (\ref{sumrule}) in terms of the 
intrinsic one-hole states is a consequence of the 
special nature of a harmonic oscillator mean-field; in contrast to 
fixed-center systems it does not hold for general mean-field s.p.\  
wave functions. As an example we can distort the h.o.\ mean field by taking 
different h.o.\ length parameters $b$ and $b'$ for the $s$ and $p$ orbitals 
in the present $s^2p$ model. The result for $S_{s}^{(d)}$ now becomes  
\begin{equation}
S_{s}^{(d)}=\frac
{3(x^2+14x+9)^{-7/2}[(x-1)^2(x+3)^2+16(x+1)(x^2+14x+9)]}
{(x+1)[(2(2x+1))^{-5/2} +16(2(x+1)(x+5))^{-5/2}]},
\end{equation} 
with $x=(\frac{b'}{b})^2$. For $x=1$ we recover the result $S_{s}^{(d)}=1$, 
i.e.\ in a pure h.o.\ mean field the spectroscopic strength of the spin-down 
neutron is fully contained in the intrinsic $s^{(d)}$ one-hole state. 
For $x\neq 1$ we find $S_{s}^{(d)}<1$; the remainder of the strength is 
contained in more complicated configurations. 
\end{appendix}


\begin{thebibliography}{999}
\bibitem{Kel}
J.J. Kelly, Adv. Nucl. Phys. {\bf 23}, 75 (1996)
\bibitem{Lap}
L. Lapikas, Nucl. Phys. {\bf A553}, 297c (1993)
\bibitem{Rad}
M. Radici, S. Boffi, S.C. Pieper and V.R. Pandharipande, 
Phys. Rev. C {\bf 50}, 3010 (1994) 
\bibitem{Pol}
A. Polls, M. Radici, S. Boffi, W.H. Dickhoff and H. M\"{u}ther, 
Phys. Rev. C {\bf 55}, 810 (1997)
\bibitem{Mut}
H. M\"{u}ther and W.H. Dickhoff, Phys. Rev. C {\bf 49}, R17 (1994)
\bibitem{Geu}
W.J.W. Geurts, K. Allaart, W.H. Dickhoff and H. M\"{u}ther, 
Phys. Rev. C {\bf 53}, 2207 (1996)
\bibitem{Pie}
S.C. Pieper, R.B. Wiringa and V.R. Pandharipande, 
Phys. Rev. C {\bf 46}, 1741 (1992)
\bibitem{asy}
D. Van Neck, L. Van Daele, Y. Dewulf and M. Waroquier, 
Phys. Rev. C {\bf 56}, 1398 (1997) 
\bibitem{Die}
A.E.L. Dieperink and T. de Forest Jr., 
Phys. Rev. C {\bf 10}, 543 (1974) 
\bibitem{future}
D. Van Neck, and M. Waroquier, in preparation.
\bibitem{del}
D. Van Neck, A.E.L. Dieperink and M. Waroquier,
Phys. Rev. C {\bf 53}, 2231 (1996)  
\bibitem{Cle}
C.F. Clement, Nucl. Phys. {\bf A213}, 469 (1973)
\bibitem{Leu}
M.B. Leuschner {\em et al.}, Phys. Rev. C {\bf 49}, 955 (1994)
\bibitem{Wir}
R.B. Wiringa, R.A. Smith and T.L. Ainsworth, 
Phys. Rev. C {\bf 29}, 1207 (1984)
\bibitem{Sch}
R. Schiavilla, V.R. Pandharipande and R.B. Wiringa, 
Nucl. Phys. {\bf A449}, 219 (1986)
\bibitem{deVries}
H. de Vries, C.W. de Jager, and C. de Vries, 
At. Data Nucl. Data Tables {\bf 36}, 495 (1987).
\bibitem{Sch2}
R. Schiavilla and V.R. Pandharipande, 
Phys. Rev. C {\bf 36}, 2221 (1987)
\bibitem{Pud}
B. S. Pudliner, V. R. Pandharipande, J. Carlson, S. C. Pieper, and
R. B. Wiringa,
Phys. Rev. C {\bf 56}, 1720 (1997).
\end{thebibliography}
\end{document}